\documentclass[graybox]{svmult}

\usepackage{mathptmx}       
\usepackage{helvet}         
\usepackage{courier}        
\usepackage{type1cm}        

\usepackage{makeidx}         
\usepackage{graphicx}        
\usepackage{multicol}        
\usepackage[bottom]{footmisc}

\makeindex             

\begin{document}
\title*{A Newtonian approach to the cosmological dark fluids}

\author{A. Aviles, J. L. Cervantes-Cota, J. Klapp, O. Luongo and H. Quevedo}

\institute{Alejandro Aviles  \at Departamento de Matem\'aticas, Cinvestav del Instituto Polit\'ecnico Nacional (IPN), 07360, M\'exico DF, M\'exico. \\
\email{aviles@ciencias.unam.mx}
\and
Jorge L. Cervantes-Cota \at Departamento de F\'{\i}sica,  Instituto Nacional  de Investigaciones
Nucleares, Apartado Postal 18-1027,  11801 M\'exico DF, M\'exico. \\
\email{jorge.cervantes@inin.gob.mx}
\and
Jaime Klapp \at Departamento de F\'isica,  Instituto Nacional  de Investigaciones
Nucleares, Apartado Postal 18-1027,  11801 M\'exico DF, M\'exico. \\
Departamento de Matem\'aticas, Cinvestav del Instituto Polit\'ecnico Nacional (IPN), 07360, M\'exico DF, M\'exico. \\
\email{jaime.klapp@inin.gob.mx}
\and
Orlando Luongo \at Instituto de Ciencias Nucleares, Universidad Nacional Aut\'onoma de M\'exico (UNAM), M\'exico, DF 04510, M\'exico. \\
Dipartimento di Fisica, Universit\`a di Napoli ``Federico II'', Via Cinthia, I-80126, Napoli, Italy. \\
Istituto Nazionale di Fisica Nucleare (INFN), Sezione di Napoli, Via Cinthia, I-80126 Napoli, Italy. \\
\email{orlando.luongo@na.infn.it}
\and Hernando Quevedo \at Instituto de Ciencias Nucleares, Universidad Nacional Aut\'onoma de M\'exico (UNAM), M\'exico, DF 04510, M\'exico. \\
Dipartimento di Fisica and ICRA, Universit\`a di Roma ``La Sapienza'',
Piazzale Aldo Moro 5, I-00185 Roma, Italy \\
\email{quevedo@nucleares.unam.mx}
}

%
\thispagestyle{empty} \maketitle \thispagestyle{empty}
\setcounter{page}{1}

\abstract{
We review the hydrodynamics of the dark sector components in Cosmology. For this purpose we use the approach of Newtonian gravitational 
instability, and thereafter we add corrections to arrive to a full relativistic description.
In Cosmology and Astrophysics, it is usual to decompose the dark sector into two species, dark matter and dark energy. We will use instead a 
unified approach by describing a single unified dark fluid with very simple assumptions, namely  the dark fluid is barotropic and
its sound speed vanishes.
}

\section{Introduction} \label{intro}

Currently, the most accepted picture for the study of our Universe as a whole is given by the so 
called $\Lambda$-Cold Dark Matter ($\Lambda$CDM) model of Cosmology. 
The first pillar of this model is Einstein's theory of General Relativity (GR) \cite{EinsteinGR}, in which the 
spacetime itself and the matter-energy fields that live in there
are related by second order partial differential equations, thus any distribution 
of matter will effectively {\it curve} the arena where these matter fields evolve. Despite this fact, many features of the 
evolution of the Universe can be well understood in the context of Newtonian gravity; qualitatively, consider the spacetime as a curved four 
dimensional manifold
with some characteristic curvature scale $l_H$, well below this scale the effects of curvature could be 
neglected and the Newtonian limit of GR becomes a good approximation to the whole, complete description. 
In the cosmic epochs that are relevant in this short review the characteristic scale is given by 
the inverse of the rate of expansion of the Universe, that is, the inverse of the Hubble factor $H$, 
thus we expect the Newtonian results to be valid up to
the length scale $c H^{-1}$, where $c$ is the speed of light. In this work we derive the relevant equations that govern the 
cosmic fluids evolution in Newtonian gravity and once we have done this 
we will add relativistic corrections in order to reach the complete set of equations.

A second pillar of the $\Lambda$CDM model is the Standard Model of Particles. The {\it known} matter fields of the Universe are essentially 
baryons\footnote{In the jargon of Cosmology we refer to any particle of the standard model that is not relativistic 
as a baryon, referring mainly to protons and neutrons. 
In contrast, in particle physics a baryon is a composite subatomic particle made up of three quarks.}, neutrinos and 
photons. These components can be approximated as fluids as long as the
mean free path of their microscopic entities are much smaller than the typical smallest macroscopic scale of the structure of interest.
In this review we will follow this approach by considering the matter fields as fluids that evolve according to hydrodynamical equations. 
This approximation is also valid for incoherent electromagnetic radiation, while coherency requires a detailed analysis of their 
distribution functions through the coupled Boltzmann and Einstein equations; for such a treatment see for example \cite{Ma94}.

It turns out that to describe the Universe we observe it does not suffice with the ingredients mentioned in the two previous paragraphs.  
Several independent cosmological probes show that nowadays the nature of about $96 \%$ of the energy content of the Universe 
is unknown to us \cite{CervantesCota:2011pn}. As far as today
all our knowledge of these components comes from their gravitational interaction with the standard matter fields, 
in this sense we refer to them as dark.
This {\it dark sector} is usually decomposed into 
dark matter and dark energy, from which the $\Lambda$CDM model inherits its name. 
The dark matter component has the property that clumps at all scales and it is responsible for the formation of 
the cosmic structures we observe, while dark energy fills the space homogeneously 
and provides a negative pressure which counteracts gravitational attraction and ultimately accelerates the Universe.
Nevertheless, we will show --as it is done in Ref.~\cite{Aviles:2011ak}-- that the dark sector can be described 
by just one dark fluid which can be characterized
with very simple assumptions, and that there is no observation relying on zero and first order cosmological perturbation theory that
can distinguish it from the $\Lambda$CDM model, concluding that the standard decomposition of the dark sector is arbitrary \cite{Kunz:2007rk}.

The paper is organized as follows, in Sec.~\ref{Section:NCB} we develop the background evolution of the Universe in Newtonian theory, 
where arguments are supplemented to understand the results in a curved relativistic framework. In Sec.~\ref{Sect:SPNC} we study
the theory of small perturbations to the background evolution, which thereafter are generalized to curved spacetimes. 
In Sec.~\ref{Sect:DF} we introduce the dark fluid and show explicitly that it is degenerated with
the $\Lambda$CDM model. Finally in Sec.~\ref{Sect:Concl} we summarize our results.

\section{Homogeneous and isotropic Cosmology in Newtonian gravity}  \label{Section:NCB}

One of the cornerstones of Modern Cosmology is that the Universe is homogeneous and isotropic at very large scales (from above about 150 Mpc), 
and this paradigm is called the Cosmological Principle. We observe essentially the same structures on the sky,
a random field of distribution, type and composition of galaxies; moreover, as we look in any direction we detect the 
same background of cosmic microwave background radiation
with a blackbody spectrum at a temperature of $2.725 \, \rm{K}$ with slight differences of the order of $10^{-5}  \, \rm{K}$. Assuming that 
we do not live in a privileged position in the Universe, the foundations of the Cosmological Principle relies on firm grounds.

To properly discuss this large scale scenario in a Newtonian framework, consider a spherical region of the space and the total mass $M$ contained in it, 
and denote the radius of that sphere by $R(t)$, which is in general a function of time. Take a small region over the sphere with mass $m$.
Ignoring all other forces except for gravity, the homogeneity of the Universe allows us to write the conservation of
energy $E$ as
\begin{equation} \label{NCB:PreFriedmanEq}
 \left( \frac{\dot{R}}{R} \right)^2 = \frac{8 \pi G}{3} \rho_M + \frac{2 \epsilon}{R^2}
\end{equation}
where we defined the  mass density $\rho_M \equiv 3M/(4\pi G R^3)$, $\epsilon$ is the 
energy per mass unit, and a dot means derivative with respect to time $t$. 
We can write $R(t) = c \tau_0 a(t)$ where $c$ is the speed of 
light, $\tau_0$ an arbitrary time scale and $a(t)$ a dimensionless
function of time called the scale factor. Then we define $K \equiv - 2 \epsilon/ c^2 R_0^2$, and we choose $\tau_0$ 
such that $K$ can take one of the three values $-1$, $0$ or $1$, and from Eq.~(\ref{NCB:PreFriedmanEq}) 
we obtain the Friedmann Equation
\begin{equation} \label{NCB:FriedmanEq}
H^2 \equiv \left( \frac{\dot{a}}{a} \right)^2 = \frac{8 \pi G}{3} \rho - \frac{K c^2}{a^2},
\end{equation}
where we have used the Einstein mass-energy relation to write the energy density $\rho = \rho_M c^2$ and redefined the time
$t\rightarrow ct$. In GR all forms of energy gravitate, therefore the use of the energy density instead of the matter density
allow us to consider other forms of energy besides matter as sources of gravity.
One interesting point is that Eq.~(\ref{NCB:FriedmanEq}) is the same as the corresponding in GR.
The root of this apparent coincidence is the equivalence principle: In GR 
Eq.~(\ref{NCB:FriedmanEq}) is obtained in a specific chosen coordinates, in these
coordinates the free fall observers have fixed space coordinates and as a consequence, 
about these observers there is a neighborhood where the laws of 
Special Relativity hold.

Note that we can solve Eq.~(\ref{NCB:FriedmanEq}) for $a(t)$ once we know $\rho(t)$, or alternatively $\rho(a)$. Then, we need at least one more
equation to close the system. Consider an adiabatic expansion of the same configuration, the thermodynamical Gibbs equation is then
$dE = - P dV$, where $P$ is the pressure of the considered fluid and $V=4\pi G R(t)^3 / 3$ the volume enclosed by the sphere. 
Giving the dependence on time $t$ we can write 
$\dot{E} = \dot{\rho} V + \rho \dot{V} = - P\dot{V}$, or\footnote{Note that background quantities such as $a$, $\rho$ and $P$ do no
not depend on the spatial coordinates, otherwise isotropy and homogeneity would break.}
\begin{equation} \label{NCB:ContEq}
 \dot{\rho} + 3 H (\rho + P) = 0,
\end{equation}
which is the continuity equation.
To finally solve the system of equations (\ref{NCB:FriedmanEq}) and (\ref{NCB:ContEq}) we need an Equation of State (EoS) that
relates the energy density with the pressure. In general this can be written as $P=P(\rho,S)$, where $S$ is the entropy of the fluid. 
But since in this scenario we
are restricted to adiabatic processes, the EoS can take the barotropic form $P = w(\rho) \rho$, where $w(\rho)$ 
is called the EoS parameter. Consider for the moment the case of constant $w$, in such a situation the 
continuity equation can be integrated to give
\begin{equation} \label{NCB:rhoawconst}
 \rho(a) = \rho_0 a^{-3(1+w)}
\end{equation}
where $\rho_0 \equiv \rho(a_0)$ and we have normalized $a_0 \equiv a(t_0) = 1$; in this work, as usually, $t_0$ denotes the present time. 
For example, the case $w=0$ corresponds to a very dilute fluid (commonly referred as {\it dust})
for which the energy density decays as the inverse of the volume, $\rho_m = \rho_{m0} a^{-3}$; 
the case $P=\rho/3$ corresponds to radiation for which the energy density decays 
as the fourth power of the scale factor, $\rho_r = \rho_{r0} a^{-4}$ 
---three powers for the dilution of the photons and one more for their redshift.

In GR the constant $K$ is related to the curvature of spacetime, and due to the assumption of homogeneity and anisotropy of space, 
there are only three possibilities that correspond to flat space which is the case of $K=0$, 
spherical space  ($K=1$) and hyperbolic space for ($K=-1$). Consider the case in which $K$ equals zero, we can insert the 
solution given by Eq.~(\ref{NCB:rhoawconst}) into Eq.~(\ref{NCB:FriedmanEq}) to obtain

\begin{equation} \label{NCB:FEwconst}
 \dot{a} = \left( \frac{8\pi G}{3} \rho_0 \right)^{1/2} a^{-3(1+w)/2 + 1},
\end{equation}
which can be integrated yielding
\begin{equation} \label{NCB:aoft}
 a(t) \propto t^{2/3(1+w)},
\end{equation}
for $w\neq -1$. The case $w=-1$ corresponds to non-evolving dark energy and the growth of the scale factor as a function of 
time $t$ becomes exponential while its energy density remains constant. 
In general, the situation is more complicated and an analytic expression for the scale factor cannot be found. This is because 
there are several fluids which must be considered, namely matter, incoherent electromagnetic radiation, 
massive neutrinos and possibly dark energy, and therefore, the 
energy density of each one of them must contribute to the Friedmann Equation. 

At this point it is convenient to introduce the redshift $z$ through $a= (1+z)^{-1}$, which is commonly used instead of the scale factor. 
Consider a Universe 
filled with matter ($m$), radiation ($r$), 
dark energy with EoS parameter $w = -1$ ($\Lambda$) and with a possible non-zero curvature, the Friedmann Equation can be written as
\begin{equation}
 H(z) =  \left(\Omega_\Lambda + \Omega_K (1+z)^2 + \Omega_M (1+z)^3 + \Omega_r (1+z)^4  \right)^{1/2},
\end{equation}
where $\Omega_i \equiv 8\pi G \rho_{i0}/3H^2_0$ for matter, radiation and dark energy and $\Omega_K \equiv - K c^2/H_0^2$, are the 
energy fractional content parameters at present time ---Note that $\sum \Omega_j = 1$.

Several independent probes of the expansion history of the Universe 
which include redshift-distance measurements of Supernovae type Ia \cite{Ri99-Pe99} 
and observations of Baryon Acoustic Oscillations \cite{Percival:2009xn} 
agree in the fact that nowadays the Universe is spatially very flat ($\Omega_K \simeq 0$) and the dominant components
to the energy content are dark energy $\Omega_{de} \simeq 0.7$ (with $w_{de} \simeq -1$) and matter $\Omega_M \simeq 0.3$ 
(with $w_{M} =0$), and additional 
tiny contributions of radiation are also present. 
The question whether all this matter can be provided by the standard model of particles arises, it turns out
that the answer is no for several reasons: 
The theory of Big Bang Nucleosynthesis \cite{Gamow,BBNReview} is very accurate in predicting the relative abundances of light nuclei of atoms,
these results are very dependent in the quantity of baryons $b$ present at that time, and to obtain the observed abundances it is necessary 
that $\Omega_b \simeq 0.04$; other constrictions to this parameter arise when one consider observations of the perturbed Universe, 
for example,  measurements of the 
anisotropies in the temperature of the Cosmic Microwave Background Radiation \cite{Ade:2013nlj} and 
large scale structure observations \cite{Reid:2009xm}, both agreeing on similar values to the above quoted 
$\Omega_b$. Moreover, analysis of virialized cosmic structures as clusters of galaxies and rotation curves in spiral galaxies show that
there is a lot of missing matter that we do not observe; for a review see \cite{Roos:2012cc}.  Therefore, the matter sector that fills the Universe
must be split into two components, $\Omega_M = \Omega_b + \Omega_{dm}$, one is the contribution of the standard model of particles, and the 
other is the dark matter, which comprises about $80 \, \%$ of the total matter and whose fundamental nature is still unknown.

We conclude that the origin of about $96 \%$ of the energy content of the Universe is unknown to us. In Sec.~\ref{Sect:DF} we will consider 
the possibility that the whole dark sector is composed by just one dark fluid. 

We end this section by rewriting the Friedmann equation 
(Eq.~(\ref{NCB:FriedmanEq})) for a flat space Universe as

\begin{equation}
H^2 = \frac{8\pi G}{3} \left( \rho_{r0}\frac{1}{a^4} + \rho_{b0}\frac{1}{a^3} + \rho_{dm0}\frac{1}{a^3} + \rho_\Lambda \right),
\end{equation}
where possible contribution of massive neutrinos were omitted. Note that the first two terms on the right hand side of the above equation
correspond to the ``light'' sector, while the last two to the dark sector.

\section{Small Perturbations in Newtonian Cosmology} \label{Sect:SPNC}

In the study of the Universe at small scales, the homogeneous and isotropic description is 
no longer valid. Strictly, this situation can only be completely confronted within the framework of GR. 
The problem to study the Universe at these scales is that all the symmetries present in the 
homogeneous and isotropic description are not present. A possible solution, the one we adopt, 
is to treat only with small departures to the background evolution. 

Thus, we want  to study the evolution of a fluid with energy density $\rho = \rho({\bf r},t)$ 
and velocity field $\dot{{\bf r}} = {\bf u}=  {\bf u}({\bf r},t)$ in 
the presence of a gravitational field $\Phi({\bf r},t)$.
The continuity, Euler and Poisson equations are
\begin{equation}
 \frac{D \rho}{D t} + \nabla_{{\bf r}}\cdot (\rho {\bf u}) = 0,
\end{equation}
\begin{equation}
 \frac{D \bf{u}}{D t} = -\frac{\nabla_{\bf{r}} P}{ \rho} - \nabla_{{\bf r}} \Phi = 0,
\end{equation}
and
\begin{equation}
\nabla_{\bf{r}}^2\Phi = 4 \pi G \rho,
\end{equation}
respectively. We have used the convective derivative $D/Dt \equiv \partial/\partial t + \bf{u} \cdot \nabla_{\bf{r}}$ which describes
the time derivative of a quantity at rest in the comoving fluid frame. Adding an equation of state $P=P(\rho,S)$ the problem is solvable
in principle, 
but in practice such a situation is intractable. The alternative studied here is to treat only small departures from the 
background description introduced in the previous section. 
To this end, let us first consider coordinates $\bf{x}$ which are comoving with the background evolution, these are defined by
\begin{equation}
 {\bf x} \equiv \frac{1}{a(t)} {\bf r},
\end{equation}
and the peculiar velocity ${\bf v} = a(t)\dot{\bf{x}}$, such that ${\bf u} = \dot{a}(t){\bf x} + {\bf v}$; that is, ${\bf v}$ 
is the velocity of the fluid with respect to the background comoving ({\it Hubble}-)flow.  By the chain rule, the derivatives transform as 
$\nabla_{\bf{r}} \rightarrow a^{-1} \nabla_{\bf{x}}$ and $(\partial/\partial t)_{\bf{r}}  \rightarrow (\partial/\partial t)_{\bf{x}} -  
H \bf{x}\cdot \nabla_{\bf{x}}$. (In what follows we will omit the subindex ${\bf x}$ from the spatial 
gradients and $\partial/\partial t$ should be understood as being taken at fixed ${\bf x}$.) 

We now consider perturbations to the quantities $\rho$, $P$ and $\Phi$,
\begin{eqnarray}
 \rho(\vec{x},t) &=& \bar{\rho}(t) (1+ \delta(\vec{x},t)) \label{Pertdelta} \\
 \delta P &=& c^2_s \delta \rho + \sigma \delta S \label{PertPress} \\
 \Phi(\vec{x},t) &=& \bar{\Phi}(t) + \phi(\vec{x},t) \label{PertPhi}
\end{eqnarray}
where a {\it bar} denotes background quantities that only depend on the time coordinate. We introduced also $c^2_s = (\partial P/ \partial \rho)_S$, 
the squared adiabatic sound speed and $\sigma \equiv (\partial P/ \partial S)_\rho$.
Note also that the perturbation to the energy density is $\delta \rho = \bar{\rho} \delta$.
In terms of the perturbed variables, the continuity, Euler and Poisson equation become

\begin{equation} \label{ContP}
 \frac{\partial \delta}{\partial t} + \frac{1}{a} \nabla \cdot \big((1+\delta) {\bf v}  \big) = 0,
\end{equation}

\begin{equation} \label{EulerP}
\frac{ \partial {\bf v}}{\partial t} + H {\bf v} + \frac{1}{a}({\bf v} \cdot \nabla){\bf v} = -\frac{1}{a}\nabla \phi -
\frac{\nabla \delta P}{ a \bar{\rho}(1+\delta)},
\end{equation}
and
\begin{equation} \label{PoissonP}
 \nabla^2 \phi = 4 \pi G a^2 \bar{\rho} \delta.
\end{equation}

The first two equations are quadratic in the perturbed variables, therefore, in the following we treat them as small
and linearize Eqs.~(\ref{ContP}) and (\ref{EulerP}) to obtain

\begin{equation} \label{ContP2}
 \frac{\partial \delta}{\partial t} + \frac{1}{a} \nabla \cdot {\bf v}  = 0,
\end{equation}

\begin{equation} \label{EulerP2}
\frac{ \partial {\bf v}}{\partial t} + H {\bf v} + \frac{1}{a}\nabla \phi +
\frac{1}{a} \nabla c^2_s \delta = 0.
\end{equation}
Note that in the last equation we used Eq.~(\ref{PertPress}) and considered adiabatic perturbations only. By appealing the conservation 
of angular moment in an expanding universe, it is expected that the divergence-free piece of the peculiar velocity should decays 
quickly with time. This can be easily seen by taking the rotational of equation (\ref{EulerP2}), arriving at $\nabla \times {\bf v} 
\propto a^{-1}$, which means that in the absence of sources of vector perturbations these modes are not relevant in first order 
perturbed cosmology, allowing us to consider only the curl-free piece of the velocity in the following discussion; moreover,
any initial large vector perturbation would break the isotropy of the background, and thus it is not compatible with the Cosmological Principle.

Now, we are in position to give a closed linear second order equation for the density contrast $\delta$, taking the partial time derivative
of Eq.~(\ref{ContP2}) and using Eqs.~(\ref{PoissonP}) and (\ref{EulerP2}) we arrive at

\begin{equation} \label{Eq2delta}
 \frac{\partial^2 \delta}{\partial t^2} + 2 H \frac{\partial \delta}{\partial t} - 
 4 \pi G \bar{\rho} \delta - \frac{c^2_s}{a^2}\nabla^2 \delta = 0.
\end{equation}
In this equation the second term acts as a {\it friction} provided by the background expansion, the third term implies gravitational
attraction, while the fourth is a pressure term, showing the important aspect of the competition between
gravitational attraction and pressure  support. 

Being the set of partial differential equations linear in the perturbations it is convenient to work instead in Fourier 
space\footnote{Our convention for a Fourier transform of a vector or a scalar function $f$ is
\begin{displaymath}
 \tilde{f}({\bf k}) = \int d^3 x f({\bf x}) e^{i {\bf k} \cdot {\bf x}}.
\end{displaymath}
In this work, without worrying about confusions, we omit the tilde on Fourier transform quantities.


}, 
arriving to ordinary differential equation for which each Fourier mode evolve independently. 

We define the variable $\theta$ as the
divergence of the velocity in Fourier space, that is

\begin{equation} \label{thetadef}
 \theta \equiv - \frac{i}{a} {\bf k} \cdot {\bf v}.
\end{equation}
The factor $a^{-1}$ is a convention used since the size of a 
perturbation $\lambda \sim k^{-1}$ grows with $a$, and thus $k/a$ becomes a comoving wave number.

In Fourier Space $\nabla \rightarrow - i {\bf k}$, and Eqs.~(\ref{PoissonP}), (\ref{ContP2}) and (\ref{EulerP2}) can be written as 

\begin{equation} \label{PoissonPF}
  k^2 \phi = - 4 \pi G a^2 \bar{\rho} \delta,
\end{equation}

\begin{equation} \label{ContPF}
 \frac{d \delta}{d t} + \theta  = 0,
\end{equation}

\begin{equation} \label{EulerPF}
\frac{d \theta}{d t} + 2H \theta - \frac{k^2}{a^2} \phi -
\frac{k^2}{a^2} c^2_s \delta = 0.
\end{equation}
To obtain the last equation we have taken the dot product of $i{\bf k}/a$ 
with the Fourier transform of Eq.~(\ref{EulerP2}) and used the definition (\ref{thetadef}), and by doing this,
we have isolated the curl-free piece of the fluid peculiar velocity.

In Fourier space the Jeans equation for an expanding Universe (Eq.~(\ref{Eq2delta})) becomes

\begin{equation} \label{JeansEqF}
 \frac{d^2 \delta}{d t^2} + 2 H \frac{d \delta}{d t} +
 \left( \frac{ k^2 }{a^2}c^2_s - 4 \pi G \bar{\rho} \right) \delta   = 0. 
\end{equation}

From this last equation it should be clear the interplay between gravitational instability  and pressure support. There exist a threshold scale, called
the Jeans length $\lambda_J = a c_s \sqrt{\pi/G \rho}$, 
for which perturbations with comoving size $L \sim a k^{-1} > \lambda_J$ grow while those with $L < \lambda_J$ oscillate and decay. 

From Eq.~(\ref{JeansEqF}) we can infer the behavior of dark matter perturbations in the different epochs of the cosmic evolution. Let us first
consider a matter dominated Universe, from the Friedmann equation and since $\rho_M \propto a^{-3}$ and $a \propto t^{2/3}$, 
it follows that $H = 2/3t$ and $4\pi G \bar{\rho} = 2/3t^2$. In this case, Eq.~(\ref{JeansEqF}) has two independent solutions 
$\delta_{dm} \propto t^{-1}$ and $\delta_{dm} \propto t^{2/3} \propto a$. The growing mode of the density contrast grows linearly with
the scale factor and from equation (\ref{PoissonPF}) it follows that the gravitational potential $\phi$ is constant in this case. 
Similar analysis for the growth of the density contrast of the dark matter in the epochs dominated by radiation and dark energy show
that in the first case the growing mode is logarithmic while for the second case does not exist, but it remains constant, suppressing the formation of 
structure. Therefore, in order for matter perturbations to grow enough to form the structures we observe today, 
it must have elapsed a sufficiently long epoch in which the
expansion of the universe was driven by matter (either baryonic or dark).


The equations for the perturbed variables developed so far are valid for non-relativistic matter fields and for scales which are 
smaller than the curvature length scale $c H^{-1}$, as discussed in the Introduction. To obtain the complete equations
it is mandatory to use the theory of General Relativity and hydrodynamics in curved spacetimes. For completeness we present here
the equations  for a collection of fluids  
with equation of state $P = w(\rho)\rho$ and that do not posses anisotropic stresses.
These are the Poisson equation
\begin{equation}
 k^2 \phi = -4 \pi G \sum \bar{\rho}_i \Delta_i,
\end{equation}
where
\begin{equation}
 \Delta_i = \delta_i + 3 H (1+w_i) \theta/k^2,
\end{equation}
the continuity equation
\begin{equation}
 \frac{d \delta}{d t} + (1+w) \left(\theta - 3 \frac{d \phi}{d t} \right) + 3 H \left( \frac{\delta P}{\delta \rho} - w \right) \delta = 0,
\end{equation}
and the Euler equation
\begin{equation}
 \frac{d \theta}{d t} +2 H (1-3w) \theta + \frac{\dot{w}}{1+w}\theta - \frac{k^2 c^2_s}{a^2 (1+w)} \delta - \frac{k^2}{a^2}\phi = 0.
\end{equation}

First we want to make note that for perturbations with wavelengths well bellow the Hubble scale, {\it i.e.} $k\gg H$, the Poisson equation reduces
to the one found in the non relativistic treatment. The presence of a velocity as source of the potential $\phi$ is expected
because in GR all forms of energy gravitate.
Moreover, for non relativistic matter $w=\dot{w} = 0$ and, as we have shown above,
$d \phi/dt =0$, recovering the Newtonian Euler and continuity equations.

Now, consider the case of dark energy with EoS parameter $w=-1$. 
At any epoch of the cosmic evolution from the continuity equation follows that the 
density contrast is a constant. This feature and the fact that its energy density remains also
constant, as shown in Sec.~\ref{Section:NCB}, implies that this fluid permeates all the space homogeneously, giving it the {\it alias} of
non-evolving dark energy. Any departure of $w=-1$ would imply that their perturbations evolve and hence possibly give rise to 
the formation of dark energy structure.

\section{The dark fluid} \label{Sect:DF}

Is it possible that the properties of dark energy and dark matter to be different manifestations of the same dark fluid?
Several unified dark models of the dark sector have appeared in the literature, the prototype of 
these is the generalized Chaplygin gas \cite{Kamenshchik:CG,Bento:CG}, 
which is defined as a barotropic fluid with EoS $P_{Chap} = -A/\rho_{Chap}^\alpha$, where the 
parameter $\alpha$ lies within the interval
$0< \alpha \leq 1$. Integrating the continuity equation (\ref{NCB:ContEq}) we obtain  
\begin{equation}
 \rho_{Chap}(a) = \left( A + \frac{B}{a^{3(1+\alpha)}} \right)^{\frac{1}{1+\alpha}},
\end{equation}
where $B$ is an integration constant. This model describes a smooth interpolation between an early phase dominated by {\it dust}, with 
$\rho \propto a^{-3}$ and an asymtotical future with $\rho = {\rm constant}$. 
The intermediate phase is well described by an EoS $P= \alpha \rho$.
The tightest constraints on the parameter  $\alpha$ come from comparisons to the observed large scale matter 
power spectrum obtaining $\alpha < 10^{-5}$ \cite{Gorini:2007},
and therefore making the model effectively indistinguishable from $\Lambda$CDM model.

Other unified models that have recently attracted the attention of the cosmological community includes scalar fields,
modifications to Einstein's theory of gravity, among others; see, for example  \cite{Aviles:2010ui,DeSantiago:2011qb,Khoury:2014tka}

We now specialize to a specific model that is totally degenerated with $\Lambda$CDM at least at zero and first order in cosmological
perturbation theory, the {\it dark fluid} which was introduced in Ref.~\cite{Hu:1998tj} 
and further studied in \cite{Aviles:2011ak}, \cite{Luongo:2011yk} and \cite{DFGTD}.

We define the dark fluid as in \cite{Aviles:2011ak}, that is, a barotropic perfect fluid with 
adiabatic speed of sound equal to zero.\footnote{Other definitions are possible. 
In \cite{Hu:1998tj} the barotropic condition is not considered but 
additional conditions on its EoS are imposed. In \cite{Luongo:2011yk} 
it is defined as an ideal gas with vanishing speed of sound.} Gravitational instability is 
driven by the competition between gravitational attraction and pressure
support. From Eq.~(\ref{JeansEqF}) it follows that the condition for vanishing sound speed 
allows perturbations of the fluid to grow at all scales,  
as cold dark matter does. For the dark fluid we can write the equation of state 
without loss of generality  as 
\begin{equation}
 P_d=w_d(\rho)\rho_d,
\end{equation}
where the subscript $d$ stands for $dark$. Giving $c^2_s = dP/ d\rho = 0$ one obtains that
\begin{equation}
 w_d = \frac{P_0}{\rho_d}, \qquad\qquad P_d=P_0,
\end{equation}
obtaining that the pressure is constant. For cold dark matter, this pressure is equal to zero, but astronomical observations allow this pressure
to be non vanishing, and in fact, it could be as large as the critical density of the Universe ($\rho_c \equiv 3H^2_0 / 8\pi G$). 
For example, a recent analysis of rotation curves in low surface brightness galaxies has shown that 
$|w_{dm}| < 10^{-6}$ at the center of the galaxies \cite{Barranco:2013wy}.
This allows us to think the pressure as a source of the cosmic accelerated expansion. 
To see how this is possible, consider the continuity equation for the background evolution
\begin{equation} \label{DarkFluidRev:AdGibbs}
 \dot{\rho}_d + 3 \frac{\dot{a}}{a} (\rho_d + P_0) = 0.
\end{equation}
This equation can be integrated to give
\begin{equation} \label{DarkFluidRev:rhodEv}
 \rho_d(a) = \frac{\rho_{d\,0}}{1+ \mathcal{K}} \left( 1 +   \frac{\mathcal{K}}{a^3} \right),
\end{equation}
where $\mathcal{K} = - (\rho_{d0} + P_0)/P_0$ is an integration constant fitted such that $\rho_{d0}$ is the value of the dark fluid energy density 
at a scale factor  $a(t_0) \equiv a_0 = 1 $.
Eq.~(\ref{DarkFluidRev:rhodEv}) is what one expects for a unified dark sector fluid, that is,
a component that decays with the third power of the scale factor plus a component that
remains constant. In order for the energy density to be positive at all times, $\mathcal{K}$ must be positive and therefore
the pressure is negative and lies in the interval $-\rho_{d0}\le P_0 \le 0$, allowing
the dark fluid to accelerate the Universe. 
Eq.~(\ref{DarkFluidRev:rhodEv}) shows that the dark fluid model gives the same phenomenology as the $\Lambda$CDM at the background cosmology. 
Its EoS parameter can be written as
\begin{equation} \label{DarkFluidRev:DFEoS}
 w_d(a) = -\frac{1}{1+\mathcal{K}a^{-3}},
\end{equation}
which should be compared to the corresponding  for
the dark sector of the standard model of Cosmology, 
\begin{equation} \label{DarkFluidRev:LDMEoS}
 w_{\Lambda + dm }(a) =  -\frac{1}{1+\frac{\Omega_{dm}}{\Omega_\Lambda}a^{-3}}.
\end{equation}
One can go back and forth between the two models with the identifications
\begin{equation} \label{DarkFluidRev:DFLCDMId}
 \mathcal{K} = \frac{\Omega_{dm}}{\Omega_\Lambda},
 \qquad \rm{and} \qquad \Omega_{d} = \Omega_{dm} + \Omega_\Lambda.
\end{equation}
The first of the two equations above  can be written in the suggestive form $P_0= - \rho_c \Omega_\Lambda$.
In Ref.~\cite{Aviles:2011ak} it has been shown explicitly that the degeneracy persist at the linear cosmological perturbed level, 
and heuristic arguments are given that point to
the degeneracy is present at all orders in perturbation theory.

We notice an important physical difference between the $\Lambda$CDM and
dark fluid models. To describe the observed late acceleration of the
Universe, in Eq.~(\ref{DarkFluidRev:LDMEoS}) it is necessary to include the cosmological constant
which is then interpreted as the vacuum energy. This identification
gives rise to the well-known cosmological constant problem, probably the
most serious inconsistency in theoretical physics. The dark fluid model offers 
a possibility to avoid this problem. In fact, in Eq.~(\ref{DarkFluidRev:DFEoS}) the term 
${\cal K}$ does not contain any cosmological information which should be associated with the 
vacuum energy, but of course it remains the problem to explain ${\cal K}$. 
This problem is  however in the arena of the microscopical theory of the dark fluid, that is not developed yet. 
The dynamics of the dark fluid naturally leads to
an accelerated universe, mimicking the exact behavior of the $\Lambda$CDM
model, without any cosmological constant. Nevertheless, the final
decision about this possibility requires a more detailed investigation.

\section{Conclusions} \label{Sect:Concl}

In this pedagogical short review we have developed the theory of cosmological perturbations at the background and linear levels within the
framework of Newtonian gravity for matter-energy fields in the fluid approximation. It is remarkable that some of the important aspects
of the cosmic evolution of the Universe can be understood without the use of Einstein's theory of General Relativity. Despite this 
fact, once we have derived the Newtonian evolution equations we proceeded to add relativistic corrections to arrive to the complete set of
equations, and special emphasis has been done in finding the solutions for the evolution of dark matter and dark energy perturbations.

Due to equivalence principle and that the dark sector components ---in its more radical definition---
only interact with the ``visible'' forms of energy through gravity, it
is arbitrary to decompose the dark sector into dark matter and dark energy. In this work we also reviewed a model, 
namely the dark fluid, that is indistinguishable
from the $\Lambda$CDM. This description results very appealing because it is based on a very simple assumption, 
that is, the speed of sound
of the dark fluid vanishes identically. We explicitly show that both models are degenerated and therefore, 
it does not exist any observation based on the background
and linear perturbed cosmology that can tell the correct description.

In addition, we noticed that the dark fluid model opens the possibility of
avoiding the cosmological constant problem because it can explain the
late acceleration of the Universe, without necessarily demanding the
presence of a cosmological constant. The problem now is to understand the value of the parameter ${\cal K}$, 
as a fundamental property of the dark fluid.

\begin{acknowledgement}
AA and JK gratefully acknowledge support from the project CONACyT-EDOMEX-2011-C01-165873 (ABACUS-CINVESTAV).
OL is supported by the European PONa3 00038F1 KM3NET (INFN) Project and
wants to thank Prof. Salvatore Capozziello for useful discussions.
HQ is supported from DGAPA-UNAM, grant No. 113514, and
CONACyT-Mexico, grant No. 166391.
\end{acknowledgement}

\bibliographystyle{aipproc}   

\end{document}